# The Moon Zoo citizen science project: Preliminary results for the Apollo 17 landing site

Roberto Bugiolacchi, Steven Bamford, Paul Tar, Neil Thacker, Ian A. Crawford, Katherine H. Joy, Peter M. Grindrod, and Chris Lintott

**ABSTRACT**

Moon Zoo is a citizen science project that utilises internet crowd-sourcing techniques. Moon Zoo users are asked to review high spatial resolution images from the Lunar Reconnaissance Orbiter Camera (LROC), onboard NASA's LRO spacecraft, and perform characterisation such as measuring impact crater sizes and identify morphological 'features of interest'. The tasks are designed to address issues in lunar science and to aid future exploration of the Moon. We have tested various methodologies and parameters therein to interrogate and reduce the Moon Zoo crater location and size dataset against a validated expert survey. We chose the Apollo 17 region as a test area since it offers a broad range of cratered terrains, including secondary-rich areas, older maria, and uplands. The assessment involved parallel testing in three key areas: (1) filtering of data to remove problematic mark-ups; (2) clustering methods of multiple notations per crater; and (3) derivation of alternative crater degradation indices, based on the statistical variability of multiple notations and the smoothness of local image structures. We compared different combinations of methods and parameters and assessed correlations between resulting crater summaries and the expert census.

We derived the optimal data reduction steps and settings of the existing Moon Zoo crater data to agree with the expert census. Further, the regolith depth and crater degradation states derived from the data are also found to be in broad agreement with other estimates for the Apollo 17 region. Our study supports the validity of this citizen science project but also recommends improvements in key elements of the data acquisition planning and production.

## 1. INTRODUCTION

The Moon is the only extra-terrestrial planetary body where the provenance of geological samples and their absolute radiometric ages are known accurately: correlations between these data and censuses of local crater populations with known surface ages have been used to determine crater production functions over time (e.g., Hartmann, 1970; Neukum et al., 2001). This has allowed the derivation of age estimates for lunar terrains where radiometrically dated samples are not in hand. This approach has also been adapted to other planetary bodies with due allowance for variations in impactor populations, fluxes and velocities in different parts of the Solar System (Hartmann, 1977; Ivanov et al., 2000).

The seemingly straightforward survey of crater features on the lunar surface is complicated by several factors, not least the discrimination between primary and secondary impacts (McEwen and Bierhaus, 2006), the effects of various forms of erosion on crater morphology, which act to soften the appearance of the crater form, and the influence of illumination in remotely-sensed images, which may serve either to hide or exaggerate topography. The combination of the human eye and brain remains unparalleled in pattern recognition of the kind required to accurately identify, characterize and quantify crater populations and the morphology of individual circular features. However, human efforts are constrained by the scale of the task, the substantial numbers of craters and time required to catalogue them; even relatively small mare regions (few hundred km$^2$) contain tens of thousands of craters.

Increases in computing power and the development of suitable pattern recognition algorithms, along with expanding catalogue of high-resolution planetary images, have spurred novel approaches to automated crater surveys (e.g., Urbach and Stepinski, 2006; Vijayan et al., 2013). Nevertheless, these techniques have yet to achieve the accuracy of a human observer. Citizen science initiatives seek to bridge the gap between the human and the computational techniques (e.g., Joy et al., 2011; Robbins et al., 2014).

Moon Zoo ([www.moonzoo.org](www.moonzoo.org)) is part of the suite of Zooniverse citizen science projects (Lintott et al., 2008, 2011), which enlist thousands of science enthusiasts around the world to carry out large-scale mapping and cataloguing of astronomical phenomena. Moon Zoo is specifically devoted to mapping features on the lunar surface and forms the basis for the work reported here.

**1.1** Aims of this study

The aim of this study is to gain a level of confidence in the Moon Zoo citizen science data to generate reliable crater size-frequency distributions across the lunar surface. We also test the validity of interpreting crater size spread among users as an index of crater erosion, and by implication, age. This work focuses on the statistical analysis of small (<500 m diameter) impact craters surveyed near the Apollo 17 landing site. This region was selected for a number of reasons: a) it is the best geologically constrained Apollo landing site; b) a wide range of NAC images at different illumination conditions were available at the time of the Moon Zoo interface design; c) its geomorphologic diversity, ranging from uplands, downslopes, old maria, regolith porosity variations, and extensive secondary craters fields; d) the 40$^{th}$ anniversary of the Apollo landing coincided with the start of this project and we used this opportunity to rekindle the public interest in Moon Zoo by focusing efforts on this region. Indeed, for a period of time (18 months) only images covering the Apollo 17 site were offered to the Moon Zoo users.

**1.2** Methodology

In order to assess the reliability of the Moon Zoo citizen science output, an expert crater survey was carried out (Section **4**). A subset was marked by three other planetary scientists for validation of the larger set. We also considered the input behavioural pattern of each Moon Zoo user in order to allocate individual 'confidence' weighting parameters (Section **5.1**). Further, we developed a new method to coalesce crater data annotations (lat., long, radius) from several non-projected, uncalibrated NAC images into single, map-projected entries (Section **5.2**).

Based on the strengths and weaknesses found we propose changes and improvements in several areas of the Moon Zoo interface (Section **7**). These recommendations are also applicable to other feature-marking citizen science projects.

**2. MOON ZOO**

One of the main advantages of Moon Zoo (Joy et al., 2011) and other planetary surface citizen science projects (e.g., "Clickworkers": Kanefsky et al., 2001; "CosmoQuestX": Robbins et al., 2012; "Be A Martian!": http://beamartian.jpl.nasa.gov/maproom#/MapMars) is that they facilitate classification of large amounts of data by breaking it down into small independent observations and then recombining the results for scientific analysis. Additionally, educational research has being carried out to identify trends in the classification behaviour and site usage of Moon Zoo users over time, to assess public understanding of lunar concepts, and to determine what motivates users to take part in this project (Prather et al., 2013).

Moon Zoo was launched in May 2010 and planned to be retired by the end of 2015. Registered users identify, classify, and measure feature shapes on the surface of the Moon using a tailored graphical interface (Fig. 1). The interface is also linked to a wide range of education and public outreach material, including a forum and blog, with contributions and moderation by experts and invited specialists (Joy et al., 2011). Users undergo a preliminary training consisting of a video tutorial (although this is not compulsory and one of the weaknesses of the current implementation is that some users may skip this key step; see Section **7** for further discussion).

**2.1** Scientific objectives

The key scientific objectives of Moon Zoo relate to the statistical population survey of small craters, boulder distributions, and cataloguing of various geomorphologic features across the lunar surface such as linear features, bright fresh craters, bench craters, etc. (see Joy et al., 2011 for details). This work focuses on the analysis of the crater survey results around the Apollo 17 region.

Crater count data can be employed to address three key science drivers:

(i) *Crater population statistics*: Deriving cumulative crater frequencies as a function of crater diameter (Hiesinger et al. 2010a, 2010b; Plescia et al. 2010; Robbins, 2014) allows for a model age estimation of mapped geologic units (e.g. Wilhelms et al. 1978; Hiesinger et al. 2000, 2003). This methodology is based upon models of crater size-frequency distributions (Hartmann 1977; Wilhelms et al. 1978; Moore et al. 1980; Neukum et al. 2001; Ivanov et al. 2000; Stöffler et al. 2006; and references therein) calibrated against returned Apollo and Luna samples.

Moon Zoo was designed to help refine the crater production function of small impact craters across different geologic terrains and shed new light on related aspects such as crater saturation as a function of surface maturity (Soderblom 1970; Schultz et al. 1977). New censuses of small crater populations at the Apollo and Luna landing sites would also help validate crater size-frequency distribution models and better constrain our understanding of the age of the lunar surface.

(ii) *Crater degradation index:* Since the centre and diameter of a given crater are noted by several users, the statistical variability of these circles' locations and sizes might be exploited to produce a crater degradation-state index, an indicator of relative age (Ross, 1968; Soderblom 1970; Soderblom and Lebofsky, 1972; Basilevsky, 1976; Moore et al., 1972; Craddock and Howard, 2000).

(iii) *Regolith depth estimations*: A layer of surficial regolith of variable thickness mantles the entire lunar surface, ranging from around 4.5 m (mare regions) to ~8.0 m (highlands) (Fa and Jin, 2010, and references therein). Small crater populations can be used to identify discontinuity in regolith layers (e.g., transition from regolith to underlying bedrock unit or a more consolidated megaregolith horizon) using the 'equilibrium diameter' method of determining regolith depth. This method exploits the process of impact gardening, whereby surfaces are disaggregated and overturned by impact processes, thus destroying the record of previous impact cratering events. The equilibrium diameter is identified when the cumulative number of craters seen on the surface is less than the number produced (Wilcox et al., 2005) and can be recognised as a change in slope in histograms of crater size-frequency (Gault 1970; Schultz et al., 1977). Regolith depths on the Moon are only accurately constrained at the small number of Apollo landing sites where seismic experiments were carried out. Understanding the thickness of the lunar regolith in different regions of the Moon helps to provide a different surface age estimate. Conversely, measuring the regolith thickness on units with well-determined ages constrains the regolith formation rate. Improved global regolith thickness maps also provide important information for future exploration of the Moon: for example, penetrator-based missions (Crawford and Smith 2008; Smith et al., 2009) require knowledge of regolith thickness to plan for landing site selection. A better understanding of regolith thickness distribution will also aid the quest of future resource exploitation on the Moon (e.g., Crawford, 2015) with implications for lunar drilling operations.

## 2.2 Catalogue Construction

### 2.2.1 Data set

The Moon Zoo project utilizes Planetary Data System (PDS) released high spatial resolution images (with associated metadata) from NASA's Lunar Reconnaissance Orbiter Camera's (LROC) Narrow Angle Camera (NAC) instrument (Jolliff et al., 2009; Lawrence et al., 2009; Robinson et al., 2010a), which has been orbiting the Moon since June 2009 (Robinson et al., 2010b). Image resolutions range between ~0.5 to ~1.5 m/px. The original Moon Zoo dataset of 150 NAC images (around 40 Gb of data) encompasses several Regions Of Interest (ROI), including the Apollo landing sites; since 2012 the project has moved into a new phase of targeted scientific objectives, focusing users' efforts on one region at a time. In December 2012 citizen scientists were asked to look exclusively at NAC images of the Taurus-Littrow region (20.2° N 30.7° E), to coincide with the $40^{th}$ anniversary of the Apollo 17 landing on the Moon. The following year, upon reaching a set output threshold, the target mission was switched again, this time to the Apollo 12 area (3.0° S 23.4° W).

### 2.2.2 Graphical interface

The Moon Zoo custom Graphical User Interface (GUI) is an Adobe Flash application based on the ActionScript programming language (Fig. 1). The Moon Zoo software Application Programming Interface and database layer were developed by the Zooniverse team at Oxford University, building on their experience with storing and analysing large amounts of citizen science data (Lintott et al., 2008).

The interface employs three different zoom levels ($\times 1$, $\times 2$, $\times 8$) to allow for both satisfactory scientific coverage and for ease of quick download times (Fig. 1). The image size presented to the user is always $600 \times 400$ screen pixels, irrespective of the zoom level. Thus, each single standard PDS NAC image file of $52224 \times 5064$ pixel size ($52224 \times 5004$ after trimming null values) produces 1278, 102, or 16 high, medium, and low zoom slices, respectively. The minimum crater size (diameter) that can be drawn at each zoom level is fixed at 20 pixels. Assuming a NAC resolution of ~1.5 m/pixel (as used in this work), this gives a minimum crater diameter of around 28, 80, and 168 m according to the zoom level, and a maximum (full) crater size of 400, 800, and 1668 m (limited by the smallest dimension of the presented image). To avoid any possible identification bias, the user is not given a scale for the presented image or offered the option of switching to an alternative magnification (zoom) version.

In the region under investigation ~92% of the presented 5,450 slices where at native resolution (i.e., no zoom), ~7% at x2, and the rest at x8. Given that only a handful of larger craters in the region requires magnified images, we find that over 99% of all marked craters in this study originates from the non-zoomed base images.

The histograms in figures 2 and 3 show the distribution of number of annotations (craters) per slice and the number of users accessing each respectively. As we can see, slices were presented to a maximum of 15 users and then removed from the pool when this number was reached. On average, each image was accessed by users around 6 times (Fig. 3).

However, the number of craters annotated per slice are widely spread, ranging from 1 to 455 (off scale), but with a low median of 10 (Fig. 2).

The Moon Zoo project was launched when only NAC EDR (Experiment Data Record) images were available in the PDS database. These images are 8-bit, translating into a full dynamic range of 256 levels. Nevertheless, when viewed with no mission-specific calibration they sometimes appear dark and with low contrast. In order to optimise the images for the purposes of measuring craters and identifying geological features it is therefore essential to perform a rescaling of the pixel levels. An asinh($x$) brightness (where $x$ is the original CCD output value) scaling function (e.g., Lupton et al., 2004) was selected among comparable approaches. Although we initially investigated linear scaling, this algorithm was preferred for retaining good contrast in regions with very different average albedo. The function behaves linearly at low values of $x$, preserving faint details, while gradually transitioning to a logarithmic scaling at large $x$, helping to prevent bright regions from saturating.

Sometimes craters may be misinterpreted in images as topographic highs (domes etc.), and vice versa, particularly when the scene is illuminated from certain directions relative to the viewer (e.g., Liu and Todd, 2004; and references therein). It has been noted that the profile of a surface is more often interpreted correctly when appearing illuminated from above (and possibly from the upper left, Mamassian and Goutcher, 2001; Stone et al., 2009). We attempt to reduce the potential confusion due to this effect through two approaches. The first is to transform the images, by mirroring horizontally and vertically as necessary, so that the apparent illumination direction (as determined by the sub solar azimuth parameter provided in the NAC image meta-data) is in the upper left quadrant. The second approach is to include a visual indicator within the GUI that shows the expected appearance of a crater and a dome in the displayed image (Fig. 1).

The low solar incidence angles that reveal the greatest surface details can also result in significant areas of the surface to be in shadow. To avoid showing the Moon Zoo participants sub-images with little or no visible surface details, we discard sub-images for which >75% of the image pixels are black.

All the image meta-data, supplemented with details of the trimming, transformations applied and geometry of the sub-images – everything that is required to map the sub-image pixel positions back to the original NAC image – are stored in a database, which is archived for reference by later analyses. This allows for reprojecting onto the lunar sphere as more accurate control nets become available.

## 3. MOON ZOO USER DATA

**3.1** Data output

All the dimension data (crater centre location point and crater diameter derived from the circle drawn over the rim) as input by citizen scientists are recorded in image space coordinates (based on position from top left corner of the image), rather than map projected lunar coordinate space. This geo-reference system was employed as the lunar coordinate control net (Archinal et al., 2006). Subsequent analysis and science exploitation of the Moon Zoo database would therefore allow crater and feature locations to be map projected using the most up to date control net of the lunar surface. The co-registration of data generated from different NAC images covering overlapping regions represents a crucial aspect of the whole project to: (1) compare annotation patterns between images obtained under different illumination conditions and (2) aggregate multiple entries for the same features using different NAC images of the same region of the lunar surface. In order to accomplish this, geometric coordinates between the NAC images, and associated generated data, must be identical. Unfortunately, for the NAC images investigated here, the spacecraft pointing calibration (SPICE kernels) which assigns absolute coordinate points to image pixels, is not consistent at the tens to hundreds of meter spatial scale. To investigate the location of the Apollo 17 lunar module (LM) was identified on 11 different images of the landing site area and the results are shown in Table 1 and Fig. 4. Using the ISIS3 "qview" GUI (Gaddis et al., 1997; Anderson et al., 2004) we located the LM on each NAC image and read off the output coordinate values. As we can see from both Table 1 and Fig. 4 the divergence with published location data (Davies and Colvin, 2000; Haase et al., 2012) are up to ~370 m N-S and ~150 m E-W. The differences appear to follow a temporal trend with the oldest NACs showing larger offsets (i.e., orbits 203-205).

Consequently, from the onset reprojecting crater data within the meter accuracy required for Moon Zoo user annotation amalgamation from several overlapping images is potentially problematic. After several attempts using different re-projecting approaches, mostly using the 'Georeferencing' tools in Esri's ArcGIS platform, we decided to concentrate our initial efforts on the Moon Zoo annotations on two NAC pairs only, M104318871 and M104311715, acquired on the same date (7$^{th}$ August 2009) and under nearly identical illumination conditions (~57°). In the longer term, but beyond the remit of this paper, we plan to develop more effective techniques to make data generated from different NAC images of the region (see Table 1) fully comparable.

All Moon Zoo crater data used in this work, including expert counts (RB), post-processing data, keys and descriptions are available at the site: data.moonzoo.org and at the online version of this paper as supplementary material.

**3.2** Moon Zoo user statistics and weighting

The advantage of citizen science projects such as Moon Zoo is their potential for attracting large numbers of volunteers. However, minimally trained individual users are not expected to generate particularly accurate crater counts, and even 'experts' may not achieve highly repeatable results (Prather et al., 2013; Robbins et al., 2014). Different motivations and level of commitment drives each citizen scientist and influences the quality of their output. For instance, the casual user who only visits the site once or twice might not apply the size tool correctly and instead produce crater annotations of the same (default) size. This factor is clearly noticeable at all zoom levels (20, 80, 160 px, Fig. 5) and it is statistically significant (around 40% more craters counted than expected following a trend line). This issue can be in part addressed either numerically (interpolation to neighbouring bins) and/or by minimising the problem by assigning weighting attributes to individual users according to their overall output volume and its proportion of minimum sizes (Simpson et al., 2012). The latter approach is the preferred one in this work. Fig. 6 summarises the data processing steps employed to analyse Moon Zoo users' contributions.

## 4. EXPERT COUNT VALIDATION

**4.1** Method

An expert count was carried out by the lead author (RB) on similar left and right NAC image pairs M104311715 for a region of ~400 km$^2$ in size, encompassing the Apollo 17 landing site and region in the Taurus-Littrow Valley (Fig. 7). This inventory produced around 2,607 craters, ranging in diameter from 29 to 1,582 m.

In order to assess the accuracy of RB's counting output, we enrolled the help of three volunteer planetary scientists in counting craters, ranging in skill from undergraduate student (S), to postgraduate (U), to senior professional (P). A subset region of 11.62 km$^2$ (Fig. 7, inset) was selected to offer a reasonable compromise in representing a typical mare area of the valley region while allowing for a manageable crater count population size.

The chosen sample area is very challenging given the heterogeneity of the local morphology, with a wide spread of crater sizes and age/degradation. Further, the Tycho crater-forming impact event around 109 Ma (Bernatowicz et al., 1977) is thought to have produced a number of secondary impact craters that affected the region's surface, by erasing, distorting, smoothing and draping over existing craters, especially the small size population. The notation of several semi-circular depressions as potential eroded craters is somewhat subjective, and this influenced the spread in the frequency distribution of medium-size craters among observers. Eroded craters also generate a spread of centres and diameters in users' notations, although here we focus on size rather than centroid positions.

The crater surveys were carried out using the same images as Moon Zoo using the ArcGIS platform and tools. The drawing shape (a circle) had no lower radius limit set, thus, eliminating the inevitable default crater size bias afflicting even expert counters (Robbins et al., 2014). Ideally, a Moon Zoo simulated environment should have been employed, but care was taken to ensure the integrity of the validation process by using identical NAC images to the Moon Zoo platform, viewed at the same resolution, and using a crater classification tool identical to that implemented on the Moon Zoo website (a circle drawn from the centre), except, again, without a minimum (default) size. However, if one takes into consideration that by the very nature of these citizen science projects, users will be employing a wide range of different media access points, e.g., laptops, desktops, tablets, etc., each with different screen resolutions and settings, also inputting data with a mouse, a trackpad, a ballpoint etc., one soon realises the difficulty in trying to simulate these variables by one user, expert or otherwise.

**4.2** Comparison between experts

RB counted 199 craters (within the subset), 'S' 261, 'P' 202, and 'U' 358 craters, giving a mean of 255 (STD 64) craters. We have also calculated Pearson Product-Moment Correlation Coefficient (Eq. 2,) against RB, of 0.96 (S), 0.95 (P), and 0.97 (U), which represent very high correlations.

Eq. 1 $$Correl(X,Y) = \frac{\sum(x-\bar{x})(y-\bar{y})}{\sqrt{\sum(x-\bar{x})^2 \sum(y-\bar{y})^2}}$$

(*X* and *Y* are the compared values).

Fig. 8A and 8B show the Cumulative Crater Frequency (CCF) and Relative *(R)* Value (where, $R = D^3[dN/dD]$, Crater analysis techniques working group, 1979) results for each counter. We also carried out a cumulative slope comparison each side of the break-of-slope bin size (90 m), producing a mean of -0.88 ± 0.18 for craters <90 m in diameter, and -2.11 ± 0.07 for larger craters. Even on such a relatively small area, with potentially only a few hundred craters, there were some differences between counters. This confirms the variability of crater identification even among experienced crater surveyors, as found by Robbins et al. (2014) and earlier studies, reporting variations between experts in the region of ±20 % (Gault 1970).

We used root-2 binned crater sizes normalised to 100 (%) to compare the relative representation of each bin per user against RB counts ('0' baseline). Fig. 9 shows small percentage variations for each bin with all but one less than ~3 %. The larger differences are represented principally in the sub 44 m bin, where the recognition of features as craters becomes more subjective.

**5. MOON ZOO DATA CALIBRATION AND ANALYSIS**

**5.1** User filtering

The 129,479 Moon Zoo crater entries based on NAC image pairs M104311715 and M104318871 were analysed. The NACs were selected based on their similarities in illumination conditions, time of acquisition, and spatial resolution. 9,321 users generated the crater entries in the area under investigation, giving a mean of around 14 annotations per crater candidate, with 73% of users marking fewer than 10 craters, and 91% fewer than 30 (Table 2). This demonstrates a low commitment rate by a large proportion of volunteers, including a ~17% fraction who only marked one crater, at least based on these four NAC images. It is not unreasonable to question the quality and validity of all the entries generated by these citizen scientists. Consequently, we set an arbitrary threshold of minimum 'experience' in number of total crater notations to 20. This resulted in the elimination of 6,781 occasional users, and in a drop of the number of crater annotations by around a 60% (to 51,597).

With the remaining crater annotations, we devised two methods in order to minimise data 'contamination' from unreliable users. The first is a simple behavioural threshold: we eliminated all crater data from those users who marked as default sizes over 50% (P50) or 25% (P25) of their total output. The combined filtering reduces the number of annotations (from pre-clustering) by 82% (P50) and 93% (P25) and the number of users to 109 (~4% of the original) and 45 (~2%) respectively.
Admittedly, this is a rather blunt approach and it could be argued that among the rejected data there lies much correct information in terms of crater location, if not size. We plan to develop a more sophisticated filtering approach in the future where we include the default size data in aid to a better centroid estimation of the coalesced multiple crater entries.

Concurrently, we investigated a weighting approach (Eq. 2) that also takes into account the overall output volume of each citizen scientist.

Eq. 2 $\quad i = \left(1 + 0.25 * asinh\left(\frac{count}{100}\right)\right) * \sqrt{1 - \frac{count\_m}{count}}$

Where, *count* is the total crater notations and *count_m* is its default crater size fraction.

These weights are used to reject poor users (with weight below a specified minimum). We adopt the median of $i$=0.71 (W07) and a 'stricter' value of $i$=0.90 (W09). The reduction in number of notations (through exclusion of selected users' output) produced is 77% (W07) and 85% (W09), and the number of users to 134 (~5%) and 70 (~3%) respectively, thus less exclusive than the blunt percentage approach.

There is no doubt that the reduction in the number of 'trusted' volunteers might seem rather drastic and maybe wasteful on the surface. In reality, even those reduced numbers (up to 134 for W07) are much higher than any feasible pool size of dedicated researchers and volunteers. Further, better analytical approaches might in the future help 'rescuing' more of the discarded contributions as mentioned earlier (by retaining positional information from default-size entries) and, for example, by bringing back data from users who contributed heavily on different Moon Zoo tasks or NAC images, thus lowering the minimum 'confidence' level (for instance, 15 instead of 20 overall entries).

**5.2** Data Clustering – methods

Our most basic requirement is that multiple entries relating to the same circular feature must coalesce into a single entry to allow further analysis: this is by no means a trivial step. While the strength of citizen-science projects such as Moon Zoo is founded upon their brute-force parallel approach to producing surveys of very large sample sets, this project has exposed some inherent weaknesses of the method. The combined lack of experience and light training produces significant variations in crater boundary estimation (diameter) and centre location between users, especially where craters rims are highly degraded or the solar illumination angle contrives to mute the apparent topographic relief. Furthermore, as highlighted earlier (section 3.2, Fig. 5), a substantial number of users applied the default (minimum) circle size tool to annotate craters close to this diameter and, more problematically, many used the minimum size marker as a crater marker, irrespective of the crater's actual diameter. Therefore, prior to application of the clustering algorithms it is necessary to filter the user data based on apparent behavioural patterns.

5.2.1 Clustering multiple entries

Two clustering methods were investigated here:

*(1) Fastcluster* ('*f-*') approach

Multiple user entries for the same circular feature need to be collapsed to a single centroid and associated mean radius. Firstly, we apply single-linkage agglomerative hierarchical clustering (also known as nearest-neighbour or friends-of-friends clustering), using a fast Python implementation called '*fastcluster*' (Müllner, 2013). This has the advantage, versus many other clustering algorithms, of not relying on a prescribed number of clusters. This suits the nature of the Moon Zoo crater format since the number of final craters cannot be estimated a priori. The clustering is performed in x, y, and diameter, with linking length selected by inspecting the results from a range of values. The diameter difference is taken to be zero where a marking has the default (minimum) crater size. These default-size markings thus contribute to defining the existence and centre of a crater, but not its size. Final cluster craters are given a score by summing the user weights of their constituent markings. Default size craters are only given a half weighting. This score roughly corresponds to the number of markings that contributed to each crater. After clustering, we can impose a minimum score for each cluster. Here, given the non-uniform number of users per viewed image subset, we have allowed for two threshold levels: minimum of two craters, or higher.

Selected results of the *fastcluster* approach are shown in Fig. 8C-D. After removing single-annotation clusters, this method produces 7,636 final craters (~15% of the original click numbers); the stricter 'three annotations or above' rule instead reduces the number of craters to 4,685 (~9%), a 60% difference between thresholds (Table 3).

(2) *Manchester ('M-')* approach

We have also developed a Likelihood based approach to clustering utilising knowledge of the measurement errors on annotated x, y and diameter parameters. We term this the Manchester ('M-') approach because it was developed at the University of Manchester (Tar et al., 2014). The implementation is very similar to a circular Hough transform, where an x, y and diameter parameter space is populated with Moon Zoo annotations, before being smoothed with a Gaussian with width proportional to the annotation errors (which were measured to be approximately 10% of crater diameter). The smoothing has the effect of coalescing closely adjacent annotations into individual peaks, whilst preserving isolated annotations as single peaks. Each peak in this space is interpreted as an individual candidate crater, with the height of each peak be proportional to the number of annotations around that location.

It is important to stress that the Manchester method does not require setting a minimum number of crater annotations. However, for comparison, when clustering non-filtered users' data a parameter was introduced in order to simulate these conditions (M-ALL_2 and M-ALL_3; see Table 4 for abbreviations).

5.2.2 Comparison

Four pre-filtering variables based on users' behaviour (two percentile and two weighted) and two clustering thresholds (number of annotations per potential crater), produced 16 combinations, as listed in Table 3. The results show a great variation in the number of 'final' craters, from an original count of 51,597 annotations. A more detailed analysis will follow (Section **6**), but here we compare simply the post filtering-clustering crater numbers with the control set (expert, RB) in terms of deviation (in percentile): the 'Manchester' method produces the closest numbers overall, with M-ALL_3, M-P50, and M-W09 within 25% variations (see third column comparing Moon Zoo data against RB count in terms of percentile difference). Only the *fastcluster* combination *f*-W07_2 comes within this range.

**5.3** Crater degradation Indexes

It is reasonable to assume that a degraded crater, with a smoothed non-distinct rim, would have less tightly bound multiple annotations and, therefore, larger variances on measured dimensions than a fresh crater. Computing a useful index then relies upon gathering sufficient annotations per crater to generate parameter variance estimates with high enough certainty so that statistically significant changes due to degradation are observable. The error on a sample variance is given by:

Eq. 3 $$SD = var(d)\sqrt{\frac{2}{n-1}}$$

where *SD* is the error as a standard deviation, *d* is a diameter of a crater and *n* is the number of annotations around the crater. For small numbers of annotations per crater this error is large, potentially limiting the applicability of this technique. As an alternative, the apparent smoothness of degraded craters might be approximated using template craters that can be smoothed using image processing in order to find a smoothing level that gives a good match. This smoothing level might then correlate (at least at the level of a rank variable) with the variance of parameters and degradation, without requiring large numbers of annotations. Both of these methods are explored in this work:

a) The *fastcluster* (*f*-) clustering method generates a statistical spread index for each coalesced crater that can be employed to classify the spread of diameter values. The assumption is that relatively larger standard deviations in crater diameters correspond to larger uncertainties, implying poorly defined degraded crater rims.

b) The Manchester (M-) approach, in contrast, measures degradation by matching a crater image template to candidate craters using varied levels of image smoothing. An average crater appearance is computed using a selection of clustered Moon Zoo craters, with mean illumination subjected to minimise effects of albedo. This template is compared to candidate craters using a normalised dot-product match score. The amount of smoothing required to achieve the best match between a crater and this template can then be correlated with degradation, as the gradual erosion of craters mimics the appearance of a smoothed image. 16 logarithmic smoothing levels, corresponding to absolute smoothing between 0.1 to 1.9 pixels, were applied.

Morphological classes have been devised in the past to describe the degradation of small lunar craters. For example, Basilevsky (1976) assigned three degradation classes between sharp (A) and smooth (C). He also offered a quantitative description in terms of relative depth and maximum angle of inner slope. In the sub-region used as a test bench here, very few craters could be classified as "very sharp sloping craters" or 'A'; therefore, we have produced a qualitative degradation class division to the expert count based on the local crater morphologies (Fig. 10). Blue (sharp) coloured circles represent craters with the best-defined rims (Basilevsky, 1976, 'AB'), through green (defined, 'B'), orange (degraded, 'BC'), and finally red (smooth, 'C'), representing the smoothest morphologies.

We implemented a quantitative reference by using elevation figures from LOLA data (Smith et al., 2010), targeting six representative craters across the degradation scales (Fig. 10a). The depth to diameter ratios (d/D) range from around 0.05 for the most eroded craters to around 0.14 for a relative fresh one. Nonetheless, there are not enough elevation data points across this area to build a comprehensive crater depths survey; therefore, still employing said craters as references, we classified craters on a qualitative basis.

We have imported our four erosional classes ('sharp to smooth') to the clustering indexes by selecting discretional threshold points until they produced broadly comparable results with the expert ground proof. Potentially, the post-clustering erosion indexes would offer, when validated on a much larger dataset, a way to assign a quantitative, instead of qualitative, interpretation to crater erosion/age. The four classes are compared in Fig. 11 in terms of fractional percentage representation (i.e., ~60% of all expert craters are classified as 'smooth' against ~70% for M-ALL3).

# 6. RESULTS AND DISCUSSION

**6.1** (i) *Crater population statistics*

Fig. 12 compares the Cumulative Crater Frequency (CCF) and "*R*-Value" of clustered data, with the RB ('expert') count. As we can see from plots 'A-a', M-ALL2 and f-ALL3 are similar to the expert count, especially in the diameter range 30-200 m. Larger craters are statistically much fewer in number, since craters <200 m represent on average ~97% of all the surveyed craters. Indeed, one of the key Moon Zoo objectives is the mapping and study of "small craters", and we shall concentrate our analysis from here on the sub-200 m craters. Related *R* plots (Fig. 12a) confirm that the closest distribution in the unfiltered set to be M-ALL2 with f-ALL3 showing marked deviations from RB in the sub-30 m crater population.

The resultant CCF curves from (user) filtered data sets are shown in Figs. 12B-b and 12C-c. Clearly, disqualifying annotations from default-size centric users improves on the statistical representation distribution of craters, and it tends to rein in the sub-30 m diameter peak. Indeed, M-W07,9 and M-P50 show an even closer overall shape to RB than M-ALL2. It is important to stress that the 'Manchester'-filtered sets were clustered free of a minimum crater threshold parameter, explaining the larger number of craters compared to the unfiltered M-ALL2,3 dataset (see also Fig. 13B).

The Relative plot (*R*-Value) for these closest results (relative to RB) shows a surprising flattening of the crater densities against RB (Fig. 12b). Since the crater survey of the Taurus-Littrow region encompasses terrains with different histories and morphologies, it is not surprising to find a non-linear CCF representation even for the RB data. Nonetheless, given that the filtering method does not alter the statistical representation of crater sizes but focuses on setting aside 'bad users' data', one may argue that, given the higher representation of crater annotations, the Moon Zoo survey might be even closer to a hypothetical 'truer' value. However, Figs. 12C-c show what happens if the data are over-filtered, leaving only those users in that scored the least default sizes. The Moon Zoo cumulative slopes appear to resemble the 'linearity' of the RB data, but at the price of loss of (crater representation) information around 2-3 orders of magnitude. As we can deduce from Fig. 5, the default size bin also contains craters that are somewhat larger but not 'worth' the effort of resizing (in the eyes of the less committed users). The result is a subtle 'depression' in cumulative counts in the next bin size up from default.

Relative crater-size representation was further investigated in 10 m bin crater sizes and scaled to 100 (i.e., each bin represented as fraction representation of 100 % in Fig. 14). This approach was aimed at highlighting the representation of crater sizes independently from annotation numbers using the same crater data as illustrated in Figs. 8C and 8D. The result supplies further evidence of the good correlation between Moon Zoo data 'M-P50' and expert (RB) against those derived with the *fastcluster* method.

When bin sizes are scaled to 100 (%) and compared to RB, we find M-07 and -09 to be similar to M-P50 (Fig. 15). There are little differences between the three sets, with deviations from base <10 % for bins <40 m and <3 % for larger diameters. The larger discrepancies originate, as mentioned earlier, by the inclusion of somewhat larger craters to the default size and it manifests as a deficit of notations in the next crater size.

A correlation comparison (eq. 1) between the clustering datasets is shown in Fig. 16. This also shows good correlation in the <200 bin crater size range, especially the 'Manchester' set (~0.98). Larger craters do not fare well in the comparison, as they represent only ~3 % of the sampling region's population and the most noticeable anomalies appear on the largest size bins. Citizen Scientists misunderstanding larger features, like Bear Mountain or the Massifs' hillside slopes as crater walls (Fig. 7), perhaps cause such discrepancies. Bright/highland terrains appear to confuse the citizen scientists and many misunderstand images with few features and high albedo as representing relatively large craters.

**6.2** (ii) *Crater degradation index*

The two degradation indexes both broadly correlate with expert degradation classifications, at least to within a relative ordering (Fig. 10). However, when these indexes are compared to each other (Fig. 11), only those derived from the Manchester clustering method ('M-P50' and 'M-ALL3') show a good correlation with the expert's.
Discrepancies between expert classifications and indexes grow at the extremes of the size spectrum where the least and most eroded craters are found. Since statistically significant error bars are difficult to derive from subjectively selected classes, we can only speculate that a 10% variation between models to be an acceptable margin of uncertainty and good agreement.

There is a correlation between erosional state and crater size, where the sharpest rims are found mostly on the sub-90 m impact craters. This is supported by a visual inspection of figure 10a, as in the 'expert' survey, but we must be mindful of the possible methodological biases produced by the Moon Zoo interface. A faint crater that occupies a few tens of pixels across will produce smaller variations in notation size and locations then a similarly eroded, but much larger excavation.

A more rigorous and statistically robust method correlating degradation indexes derived from Moon Zoo data with classified expert categories would require calibration on much larger sample population. A key requisite would include the production of a more objective 'expert' (comparative) erosional scale, possibly developed from depth/diameter (d/D) ratio surveys derived from LRO NAC DEMs (Digital Elevation Model, Stopar et al., 2010). However, we conclude that this method shows great potential and for future work.

**6.3** (iii) *Regolith depth estimations*

Shoemaker et al., (1969) reasoned that given that regolith on a planetary surface is the product of impact events, there should be a correlation between crater Cumulative Crater Frequency (CCF) and its depth. The point that the local crater population reaches the equilibrium crater population (on the Moon a '*-2 log-log*' slope) should give us an average regolith depth. Further, he concluded that the excavation depth of an impact less the rim's height should relate to the local depth of the regolith (also, Wilcox et al., 2005).

Expert counts (RB, Fig. 8A) suggest a pivot point around the 90 m diameter bin, where the log-log slope steepens from around -0.9 to -2.0, which corresponds to the lunar equilibrium population. Crucially, the closest Moon Zoo estimate ('M-P50') compares favourably with the experts' slopes (Fig. 8C). The smaller craters sizes translate to a rough estimate of <8 m crater excavations in this region (Melosh, 1989). A four-meter regolith layer (Cooper et al., 1974; Wolfe et al., 1975) of finely comminuted mix of basalts, highland breccia and glasses represents the top layer up to the Taurus-Littrow valley. Nonetheless, the area under investigation is characterised by the northern cluster of relatively large craters in the 500-600 m size range (i.e., craters Henry, Shakespeare, and Cochise) that have deposited up to five meters of ejecta on the surface (Lucchitta and Sanchez, 1975). These would agree with the Moon Zoo-derived local regolith depth of around eight meters as derived from the break of slope.

## 7. LESSONS LEARNED AND RECCOMANDATIONS FOR FUTURE PROJECTS

This work has identified several weakness of the present Moon Zoo approach for deriving the location and sizes of crater populations. These include:

*7.1 Fixing a minimum crater size:* the provision of a minimum crater size for the crater sizing tool in the GUI, which many users just defaulted to. A significant fraction of users, probably as a combination of insufficient training and attitude, tend to rely on the default circle size to count small craters (relatively to the zoom level). For instance, on NAC images with 1.42 m/px resolution, this has produced a large peak of ~28 m craters boosted by somewhat smaller size-craters, but also affected the survey of neighbouring size (larger) craters by incorporation. This is highlighted in Fig. 15 by the underrepresentation of 30 to 40 m crater sizes. This problem should be minimised at source, by implementation of a different type of tool that does not allow for a marking unless is re-sized. This might lead to some skew or imprecision around the default size itself, i.e., in this instance few or no 28.6 m craters, but the negative effects could be minimised by the application of appropriate bin sizes.

A fully free draw size tool is not a feasible option, since much of the lunar surface is saturated at small scales by impacts and their survey would be both impractical and probably superfluous. Robbins et al., (2014) encountered similar problems with their own approach to minimum default size (a red circle represents a null value, unless resized up); consequently he discounted craters within ~4 px of the minimum diameter as not reliable. Clearly, a compromise has to be reached between ease of use, minimising the risk of user drop out, although retaining an acceptable bulk output expectation.

*7.2 User training prior to crater classification:* An improved training path should also be implemented. Standard practice in other Zooniverse projects is to provide tutorials to guide the participant through their first attempt at a task (and to be available for later reference). This is now standard in Zooniverse projects and should be implemented in future MoonZoo-like projects. That said, there might be a case for providing optional in-depth tutorials aimed primarily at assisting with more complex tasks (e.g. of recognizing complex morphological surface features). Borrowing best practices from other Zooniverse projects, future projects could also test the quality of the citizen scientists' entries in the background by randomly presenting images with known markings, thus generating a dynamic confidence index per each user (e.g. Lintott et al., 2008; Land et al., 2008; Schwamb et al., 2012; Marshall et al., 2016).

*7.3 Map projection:* On a general recommendation, the next generation of Moon Zoo should use, when possible, projected NAC or WAC images thus avoiding confusion on absolute and comparative spatial features' location. Ideally, all NAC images of the same region should conform to a base projection prior of being made accessible to the public. In this way, more accurate projection kernels could be applied when becoming available.

*7.4 Improved error analysis*: For final quantitative use, size frequency distributions require clear error bars. Poisson errors are conventionally assumed on crater counts, yet the contamination from erroneously annotated craters and craters that have been missed boost the errors above Poisson levels. A better understanding of crater counting uncertainties is needed to prevent over-interpretation (also see Robbins et al., 2014).

*7.5 Boost users' retention*: 91% of all users marked fewer than 30 craters (Table 2). A reward system, linked with better training (point *7.2* above), may motivate the volunteer to persevere towards a reward point, in terms of both quantity and quality of notations.
Eveleigh et al., 2014a - http://discovery.ucl.ac.uk/1412171/1/p79-eveleigh.pdf - explored the connection between citizen science projects and the gamification approach proposing four design considerations for "motivating and sustaining participation through gamification in citizen science".
Further, in Eveleigh et al., 2014b - http://discovery.ucl.ac.uk/1418573/1/p2985-eveleigh.pdf - they warned on the dangers of preferentially harvesting data from a few committed users against those who are just 'dabbling'. Since the outreach aspect is key to citizen science projects a balance would have to be achieved between these two sources without alienating one at the cost of the other.
These type of studies are fundamental in the understanding and development of this new frontier in big data analysis, along with large-scale projects pioneers, such as Moon Zoo.

# 8. CONCLUSIONS

In this study, we have assessed the validity of citizen-science derived data applied to the survey of craters on the lunar surface. In order to convert raw statistical annotations to science products we have developed a data management and control path from the raw Moon Zoo annotations to final crater statistics, as summarised in Fig. 6.

In summary we have: (1) validated a standard expert crater count dataset to use as comparison with Moon Zoo (Section **4**); (2) tested filtering of spoilers/bad data based on users' behavioural pattern in relation to crater default size annotations (Section **5.1**); (3) compared two different mathematical approaches in clustering multiple crater entries, one developed specifically for the Moon Zoo project (Section **5.2**); (4) derived and compared crater degradation indexes based on the spread of annotation parameters and smoothing of imaging (Section **5.3**); (5) investigated derived regolith depth from crater count frequency slopes, and found to be consistent with seismic data (Section **6.3**).

Correlations between the spread of annotation parameters, the required smoothing of images for optimal crater template matching and expertly defined levels of crater degradation show that it is in principle possible to devise an objective erosion index. The image smoothing method (Manchester, 'M-') appears to provide the highest correlation with experts and has the advantage of not requiring many annotations around each crater. However, the statistical limitations of the parameter variability approach, i.e., too few annotations per crater to give accurate erosion levels, might be extended to provide at least regional erosion levels. In this way, a regional approach could combine the spread of parameters from many local craters, boosting the accuracy of the estimated parameter variances.

We conclude that the custom built crater clustering and analysis approach ('Manchester', 'M-', Tar et al., 2014) holds much promise including a derivation of relative erosion classes, especially when combined with prior exclusion by filtering of less reliable users' data. In particular the combination of 'M-' clustering with a threshold of 50% as maximum default crater sizes notation per user (M-P50, and similar M-W07 and M-W09) produces the closest correlation with expert data (Figs. 8, and Figs. 12 to 16).

Moon Zoo produces useful science data, meeting its science objectives (Section **2.1**). However, key technical improvements in the interface are necessary to minimise users' errors. In detail: (a) change to a crater marker tool that diminishes the dangers of an overrepresentation of default (minimum) polygon size; (b) the use of projected LROC images with embedded SPICE information; (c) a more rigorous initial user training (Section **7**). In addition, a comprehensive error analysis is required to place error bars on SFD bins, as the conventional Poisson assumption does not account for extra variability caused by ambiguities in crater identifications. These issues will be the focus of future work.

# Acknowledgements

We would like to thank the tens of thousands of volunteers, the Moon Zoo citizen scientists, for their contribution to the project. We thank Arfon Smith, Robert Simpson, Anthony Cook, and Dominic Fortes for helpful conversations, and Jules Wilkinson and Geoff Roynon, moderators of the Moon Zoo website, for their patience and commitment to the cause. The manuscript has benefited from constructive reviews by Meg Schwamb and M. Ramy El-Maarry. This research has made use of the USGS Integrated Software for Imagers and Spectrometers (ISIS), and we thank the Leverhulme Trust for financial support (grant number RPG-166); KJ acknowledges funding from STFC grant ST/M001253/1.

**Supplementary information:** data.moonzoo.org

**FIGURES**

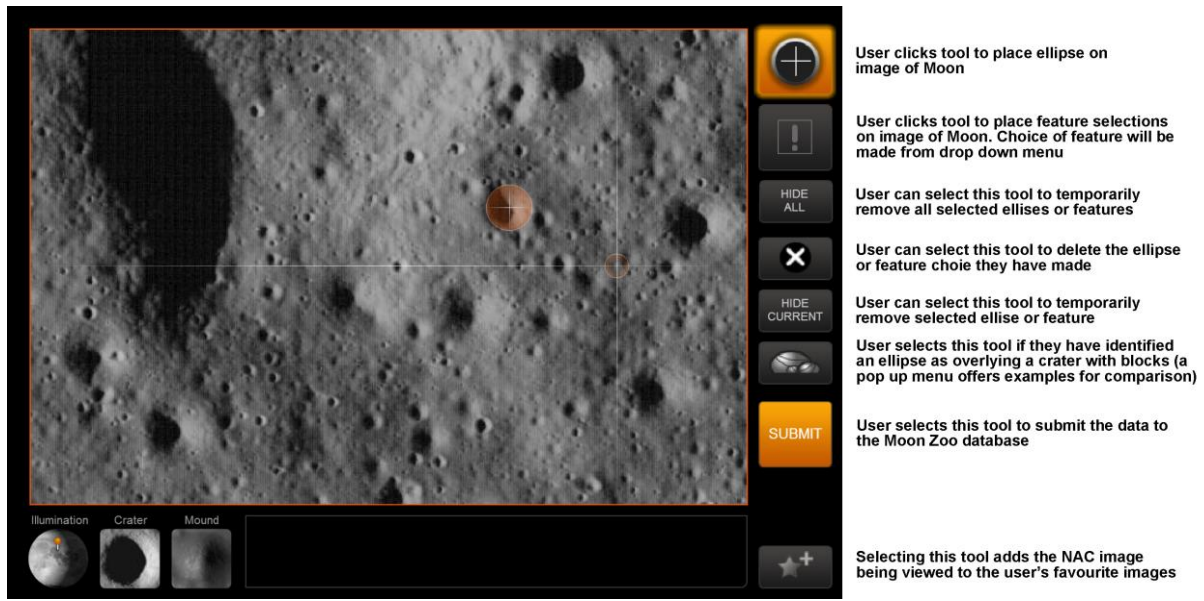

**Figure 1**. The Moon Zoo GUI (Graphical User Interface) allows users to mark craters using the aim tool, including the option of confirming or modifying the selection before submitting. Other tools help to report craters with boulders and highlight any 'interesting' features. The key in the bottom left hand corner indicates to the user what feature is a mound and which is a crater based on shadow direction.

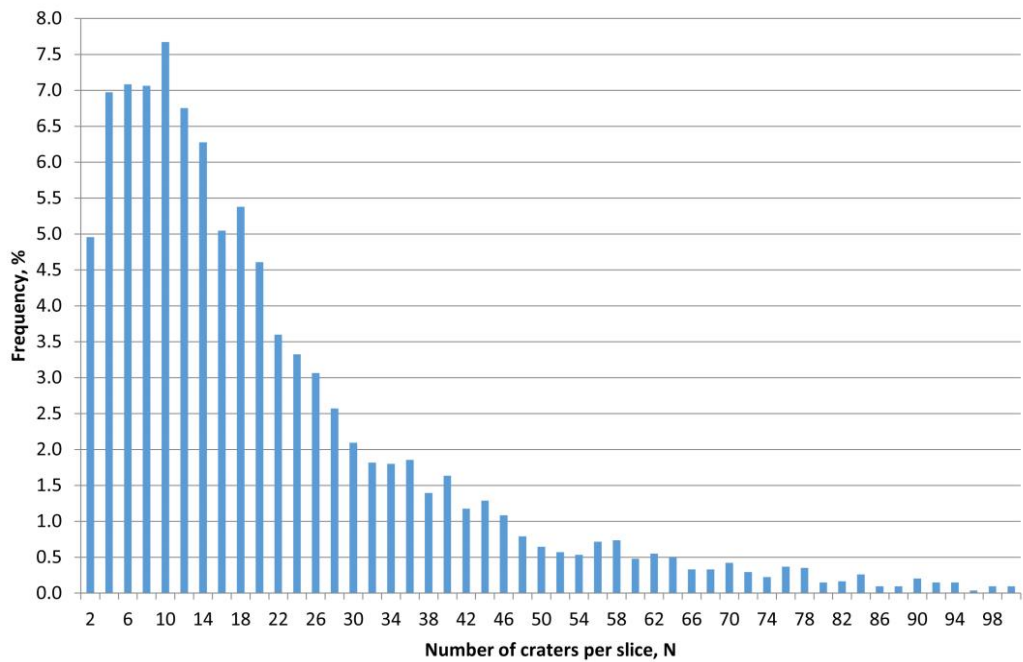

**Figure 2.** Distribution of number of craters per slice (e.g., sub-image) expressed as percentile of the total. The median is 10 craters, with a max of 455 (off scale).

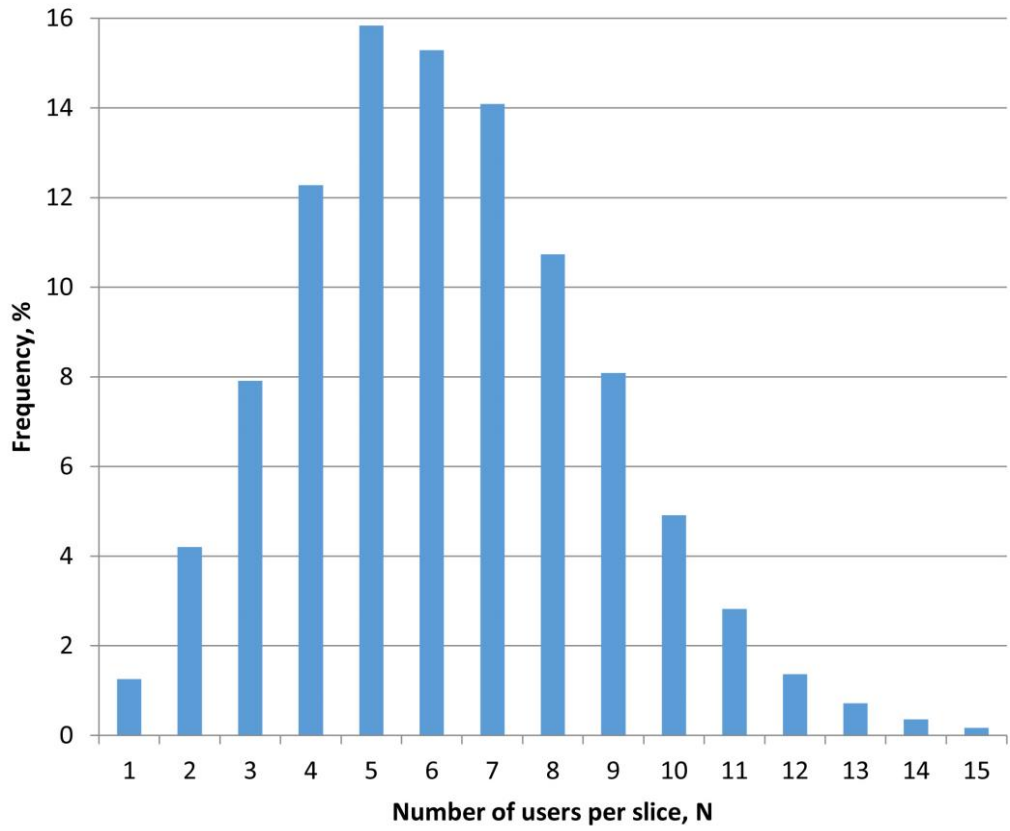

**Figure 3.** Distribution of number of users per slice (sub-image). Each slice was presented for a maximum of 15 users, with a median of 5.

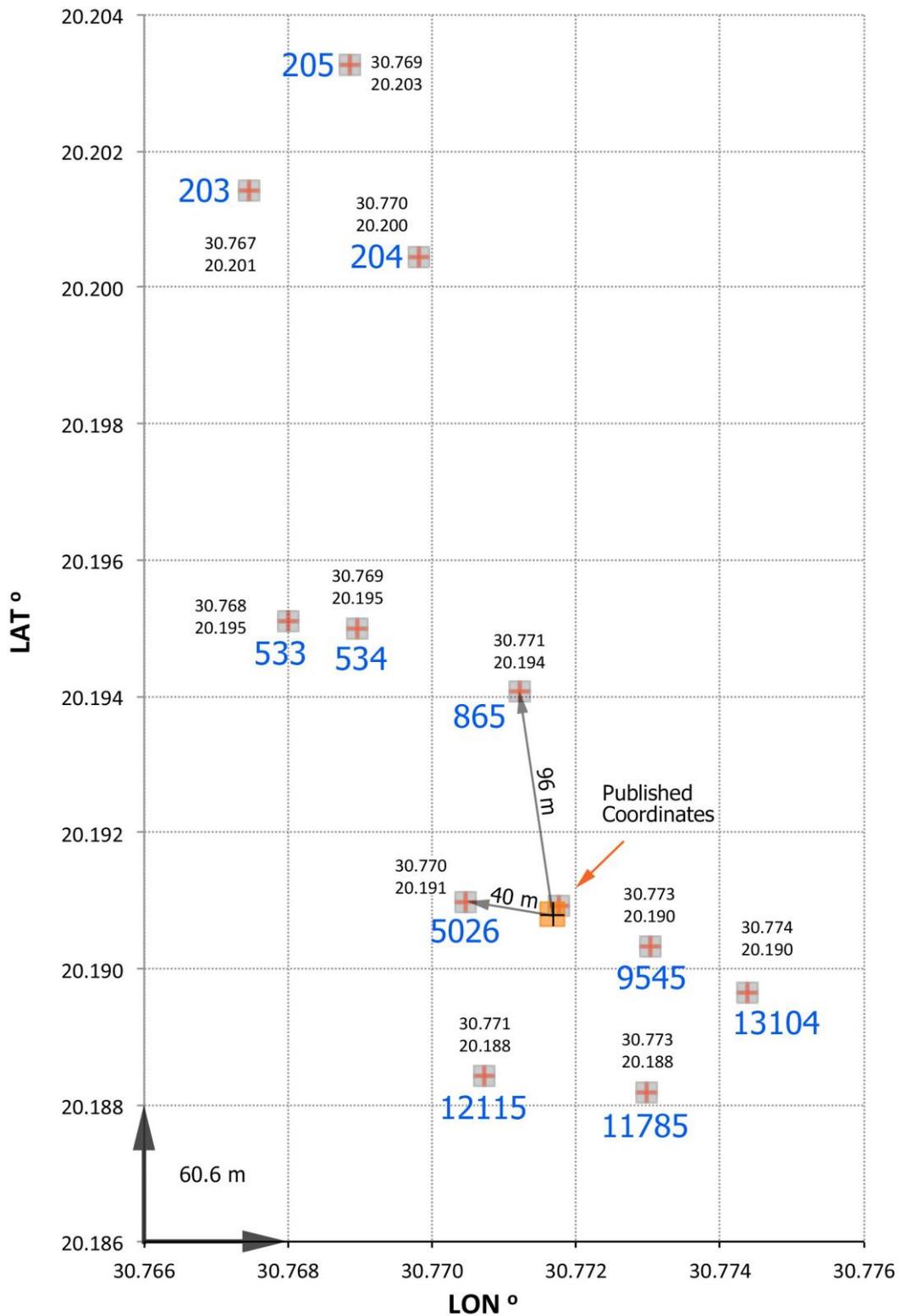

**Figure 4.** Nominal coordinates of the Apollo 17 Lunar Module as read from NAC images SPICE (Table 1). Orbit numbers are displayed close to coordinates markings. Sample distances represent ground space deviation from published landing site coordinates (Haase et al., 2012).

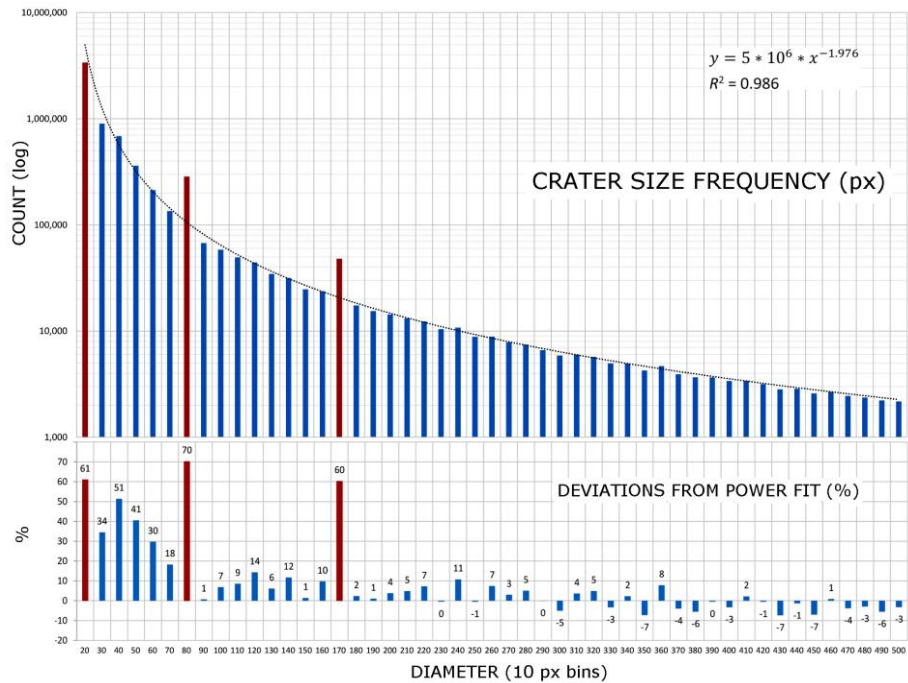

**Figure 5**. Histogram showing the number of crater annotations per pixel bin (log scale) of the entire non-clustered Moon Zoo dataset (~8 * $10^6$ annotations). The three spikes on the lower diameter range correspond to the minimum (default) diameter size on each magnification level (for the NACs used in this work, ~ 29, 110, 240 m). Inset below highlights the deviation from the interpolated fit (*y*) in terms of crater number representation, expressed as percentage.

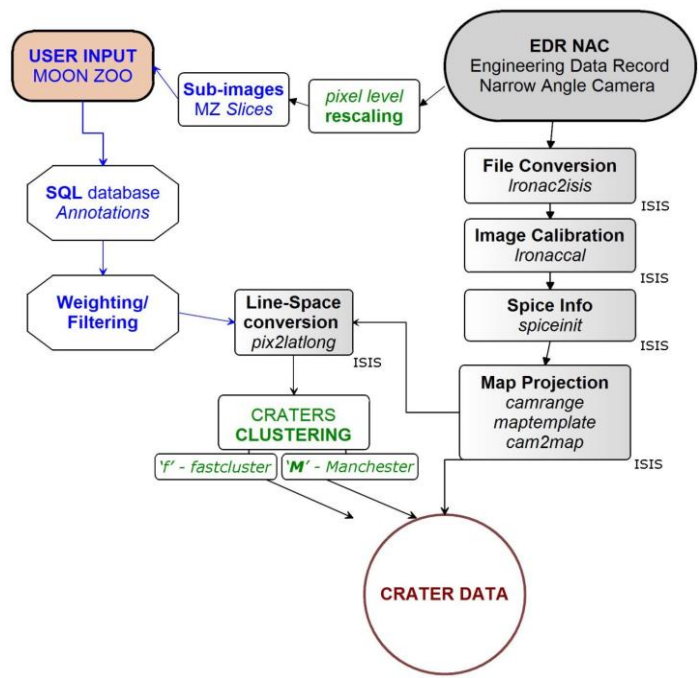

**Figure 6**. Schematics of the data proceeding steps from NAC EDR images and raw Moon Zoo data to mapped crater annotations. This paper describes in detail the journey from "SQL database - Annotations" from Moon Zoo users to (final) clustered crater data. ISIS stands for Integrated Software for Imagers and Spectrometers, USGS (Torson and Becker 1997).

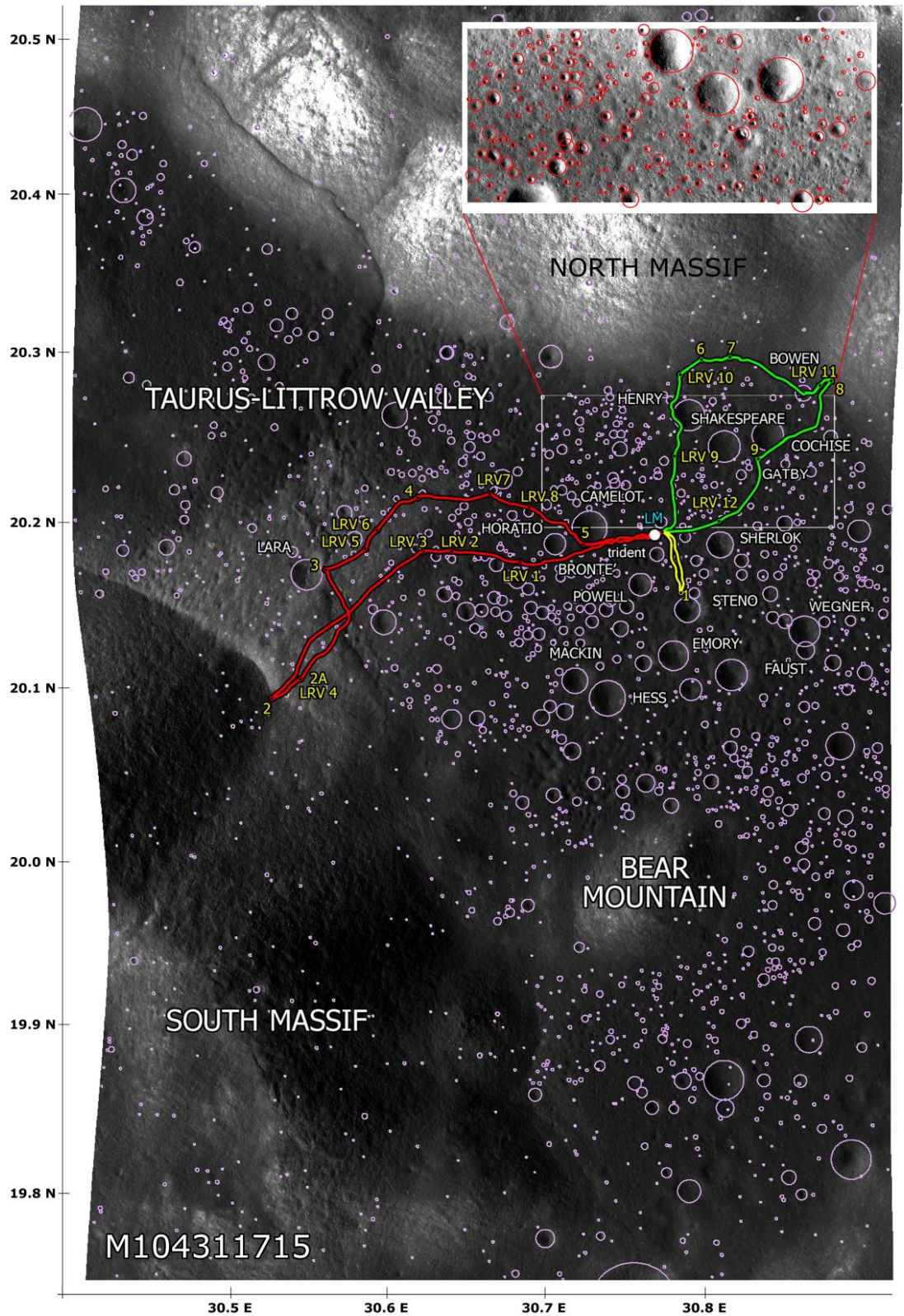

**Figure 7.** Crater annotations of expert RB. Superimposed Apollo 17 landing site and exploration map. Inset shows area used for crater survey comparison and erosion analysis. Basemap is LROC NAC M104311715 pair composite (image: NASA/GSFC/ASU).

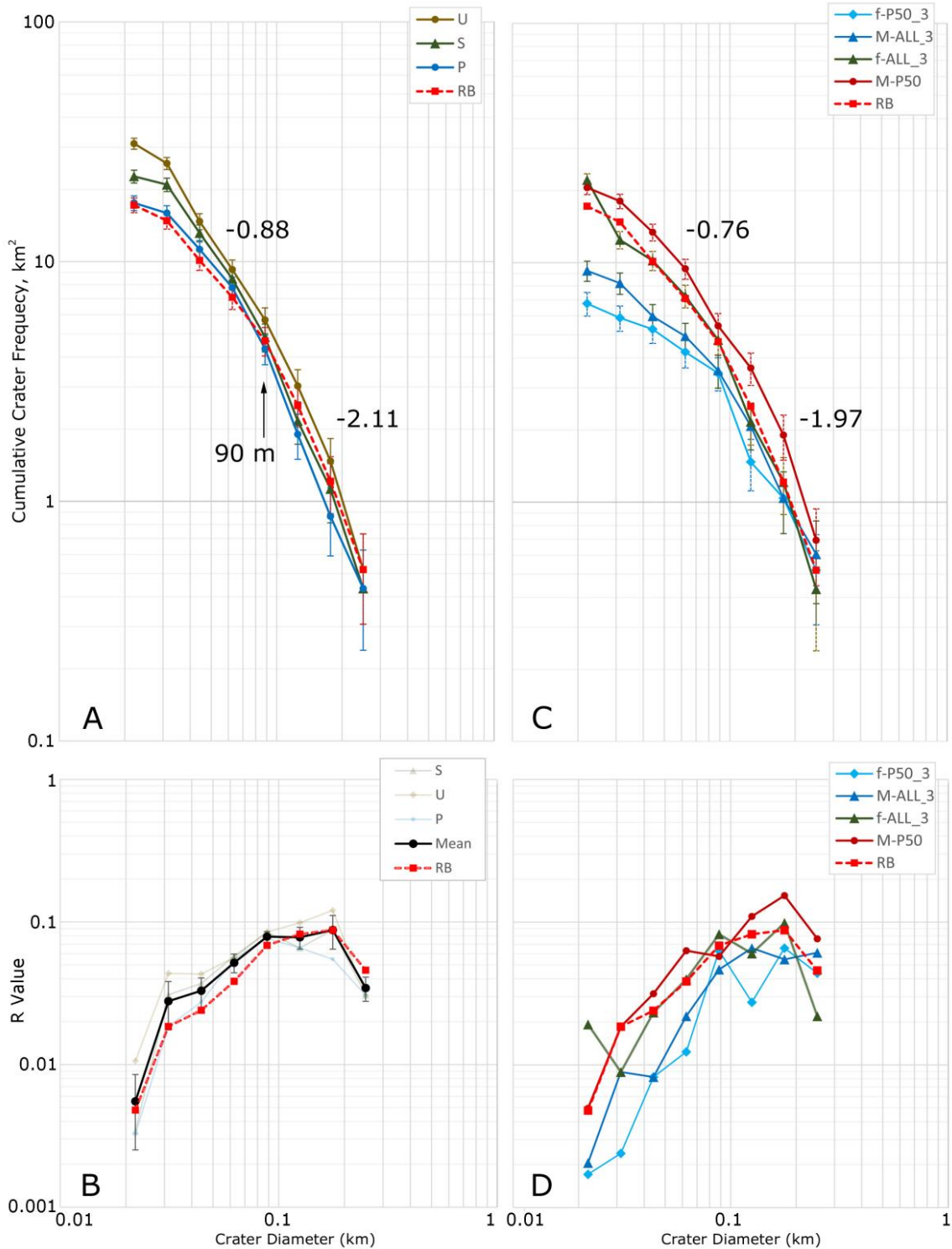

**Figure 8.** *'A':* Cumulative Crater Frequency (CCF) plot results from four crater analysts. Root-2 binned crater sizes. '**B**': same data as '**A**', relative distribution (*R*-plot) representation of deviation of the size-frequency distribution from the power law: $R = D^3$ (d$N$/d$D$). 'Mean' represents the average of the four sets. '**C**' and '**D**' similarly compere selected Moon Zoo crater results against RB. Values of the slopes are shown for crater populations with sizes <90 m and >90 m, where in '**A**' it refers to RB data and in '**C**', (MoonZoo) 'M-P50' (see Ref. Table 4 for key to abbreviations). **S**: undergraduate student; **U**: postgraduate; and **P**: senior professional.

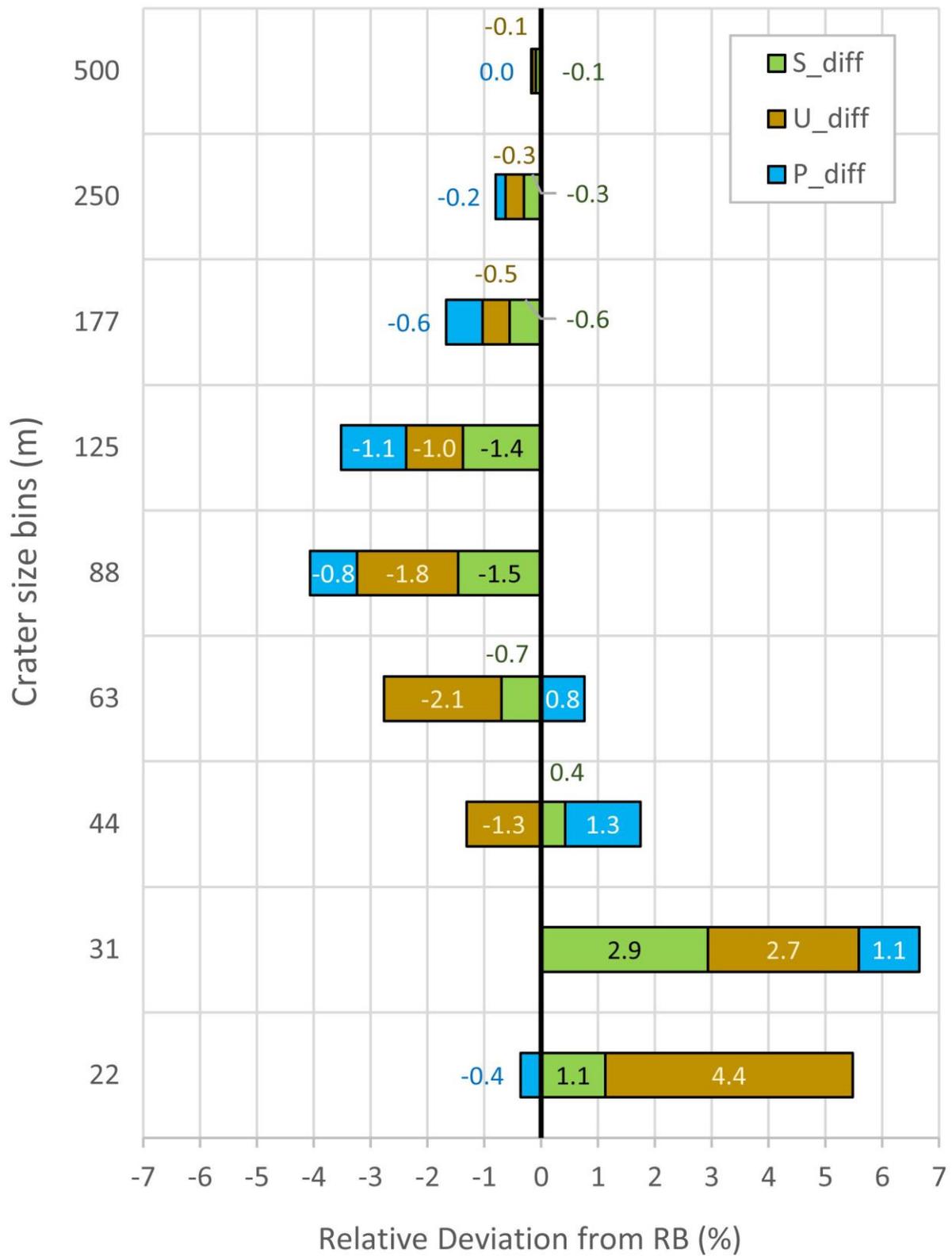

**Figure 9**. Deviations expressed as percentage between each bin and crater analysts' output **S**: undergraduate student; **U**: postgraduate; and **P**: senior professional), normalised to RB (0) for the calibration area. Root-2 binned crater sizes.

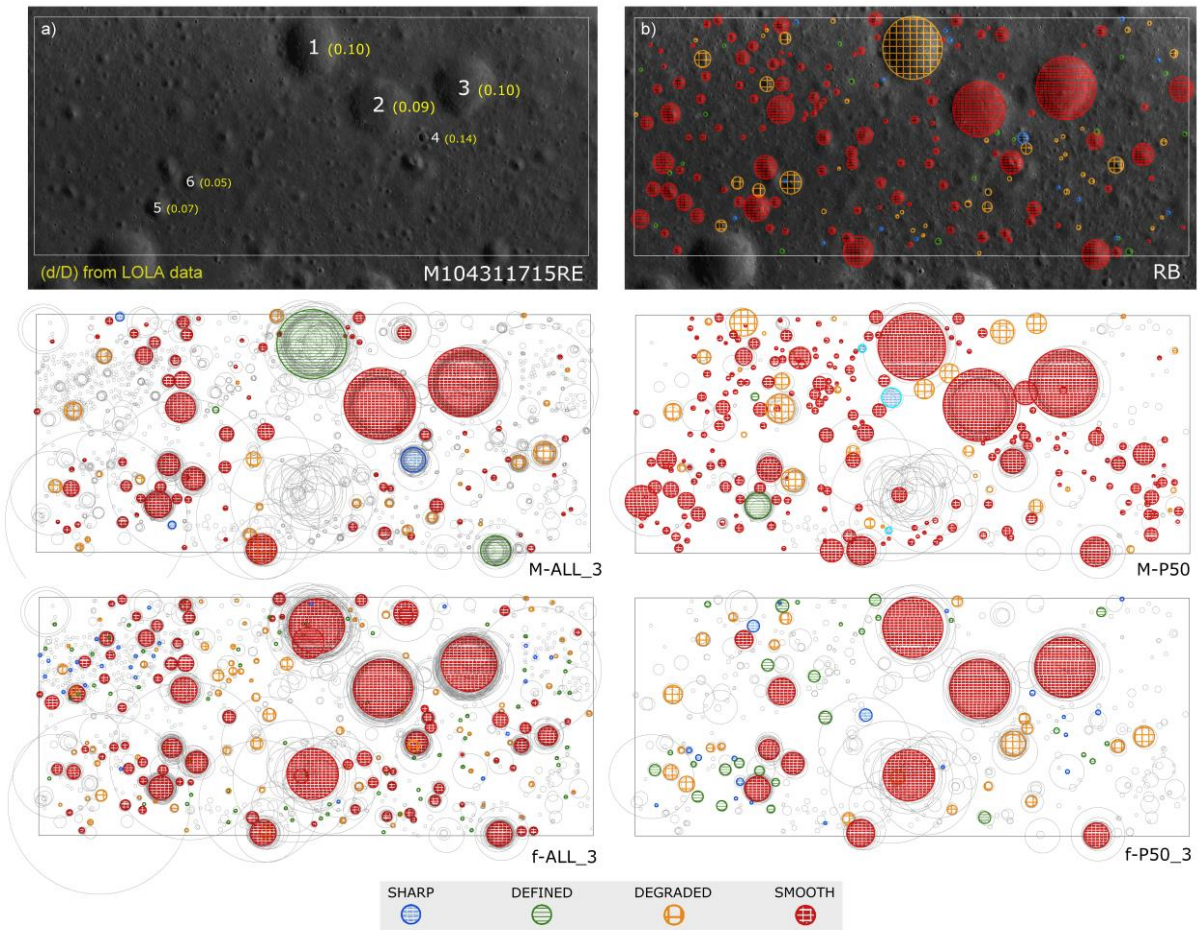

**Figure 10.** Highlighted in **a)** are six craters which depth/Diameter (d/D) ratios derived from LOLA altimetry data (Smith et al., 2010). **b)** expert crater survey is mapped including a qualitative classification of erosion state, where: blue=sharp, green=defined, orange=degraded, and red=smooth. Below, comparison of sample MoonZoo data, using the same colour code as the expert (RB). Indexes derived from the clustering 'smoothness' parameter ('M-') and STD crater size deviation ('f-'). Backgrounds show the pre-clustering Moon Zoo annotations. Ref. Table 4 for key to abbreviations.

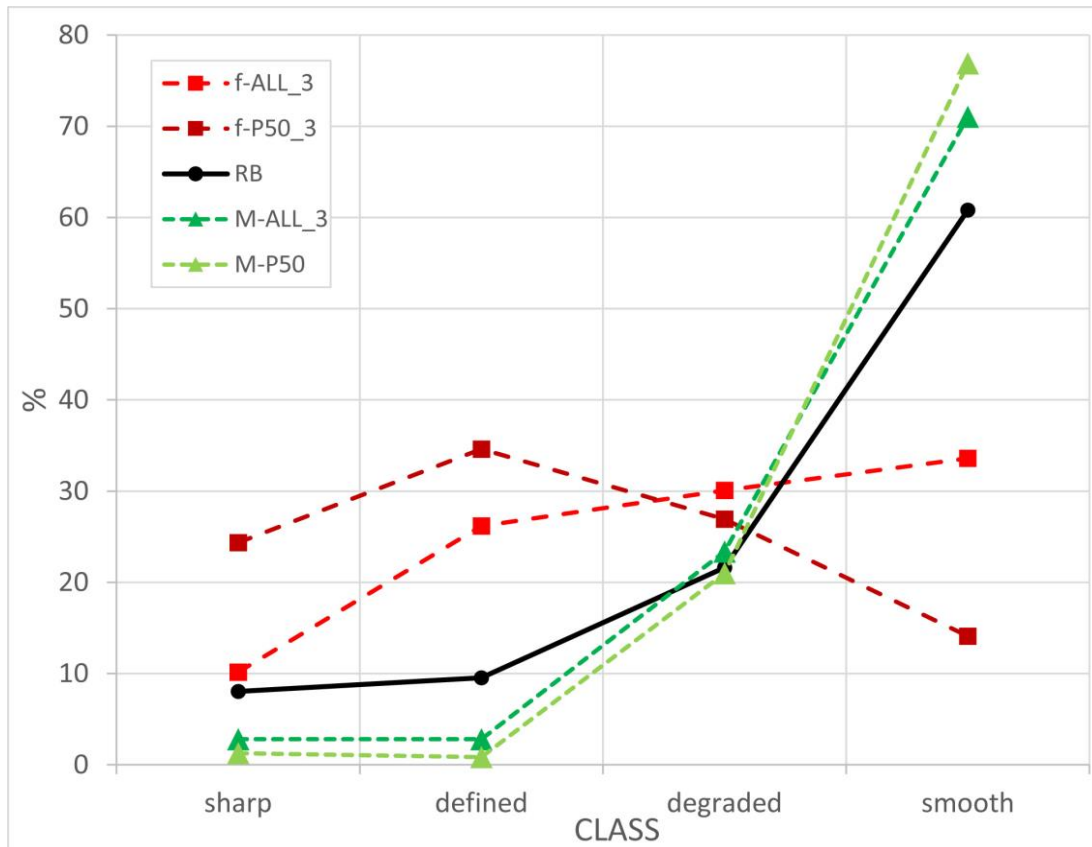

**Figure 11.** The 'smoothness' % parameter (see text) derived by the 'M-' clustering compares favourably with the qualitative erosional classes based on 'expert-RB' classification, as per Fig. 10. Uncertainties, expressed as error bars, are absent due to the subjective classification of erosional states by the expert (RB), and choices of class thresholds.

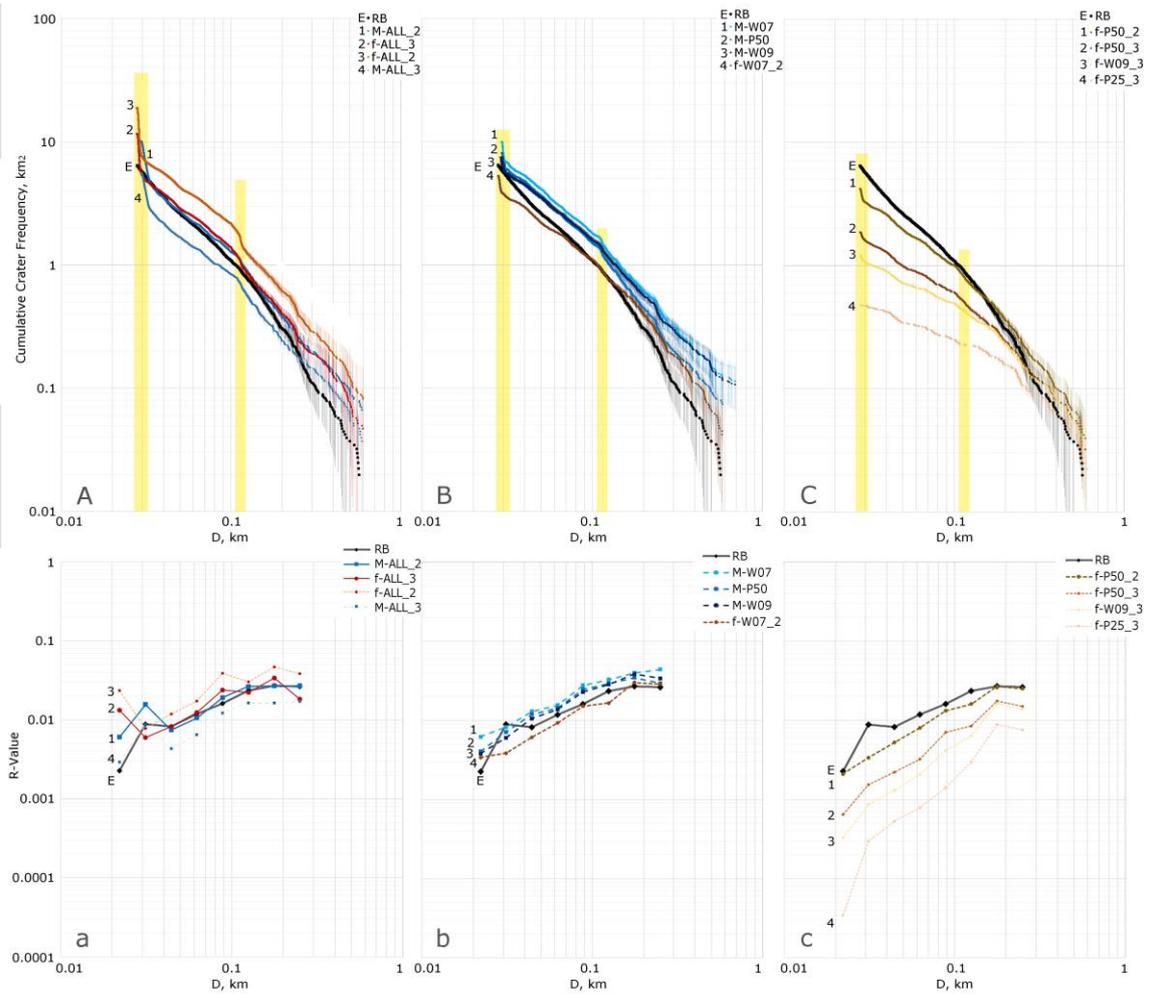

**Figure 12.** Cumulative Crater Frequency and R-values comparisons. '**A**', CCF without user's behaviour filtering. '**B**' and '**C**' compare different filtering levels against expert count, where: '**f**'=*fastcluster*; '**M**-'=Manchester; **P**=percentage; 25,50: quality threshold [by eliminating users' annotations if they are >50 (or 25)% default size (20, 80, 160 px)]; **W**=weight, a weighting algorithm that weights the default size quota and the overall number of annotations (e.g., W09 is 'stricter' than W07). Yellow bars highlight default size effects (for 20 and 80 px). See Table 4 for abbreviations used in this figure.

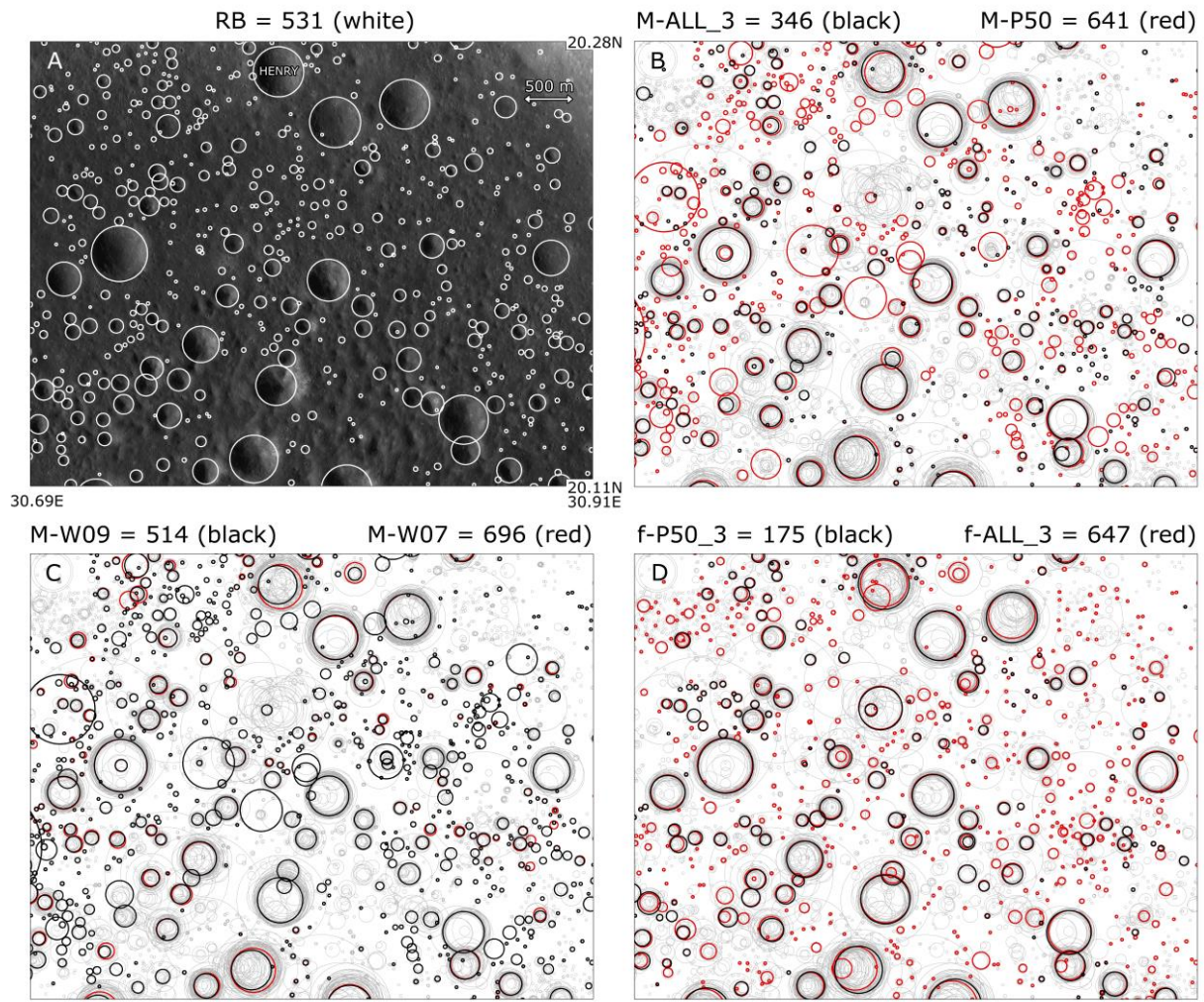

**Figure 13.** Subset (31 km$^2$) of region under study (as shown in full in Fig. 7) to illustrate the effects of different filtering and clustering approaches on Moon Zoo data, against expert (RB) count ('**A**', 531 craters) and superimposed to raw pre-clustering annotations (7,331 grey circles in the background). 'M-' (Manchester method) data is shown in **'B-C'** whereas 'fastcluster' method ('f-') is in '**D**'. For all figures ('**B-D**') black circles correspond to underlying 'red' entries thus representing a smaller subset. Full key of abbreviations in Table 4.

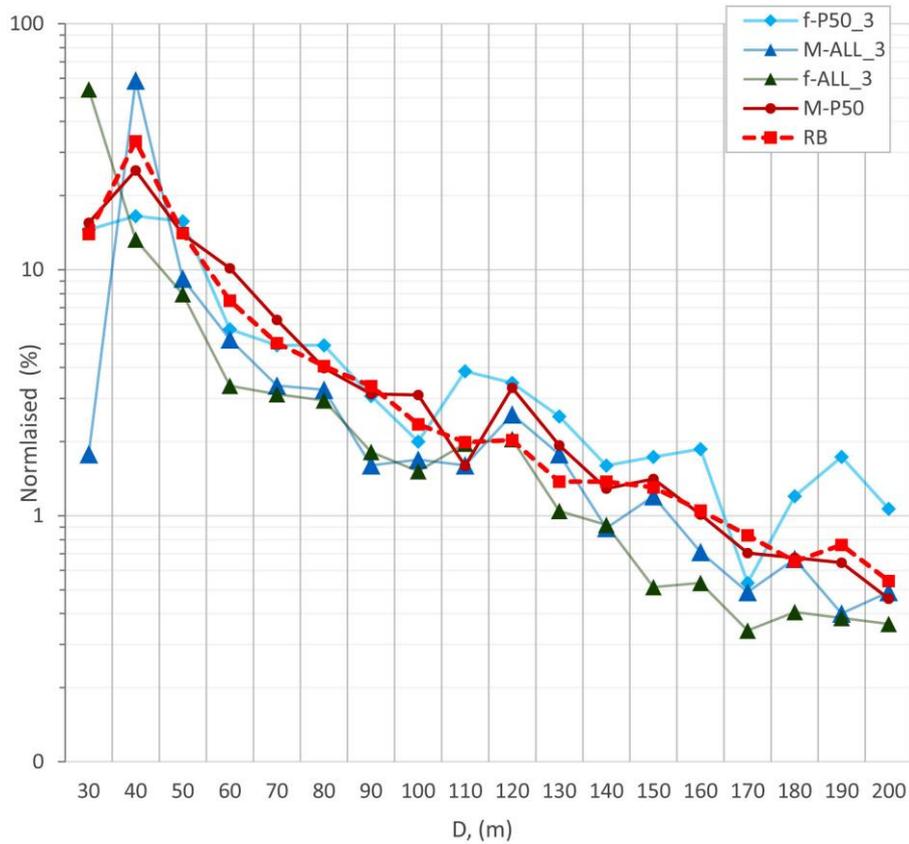

**Figure 14.** Comparison between crater bin size (10 m) representation, scaled to 100 (%) (log scale). Ref. Table 4 for key.

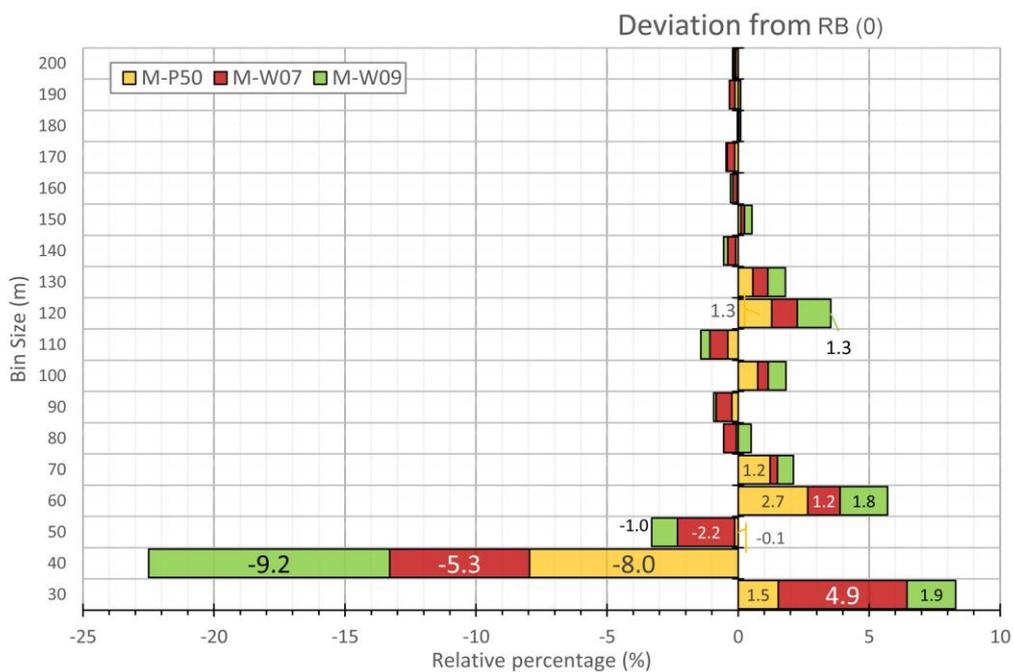

**Figure 15.** Normalised comparison, scaled to 100 (%) between chosen Moon Zoo datasets and RB (as per Fig. 14). Here the 'M-group' shows good agreement with expert count, except around the default crater sizes. The underrepresentation of craters in the 40 m bin size can be explained by users' behaviour to include these craters in the similar, default, size (~30 m).

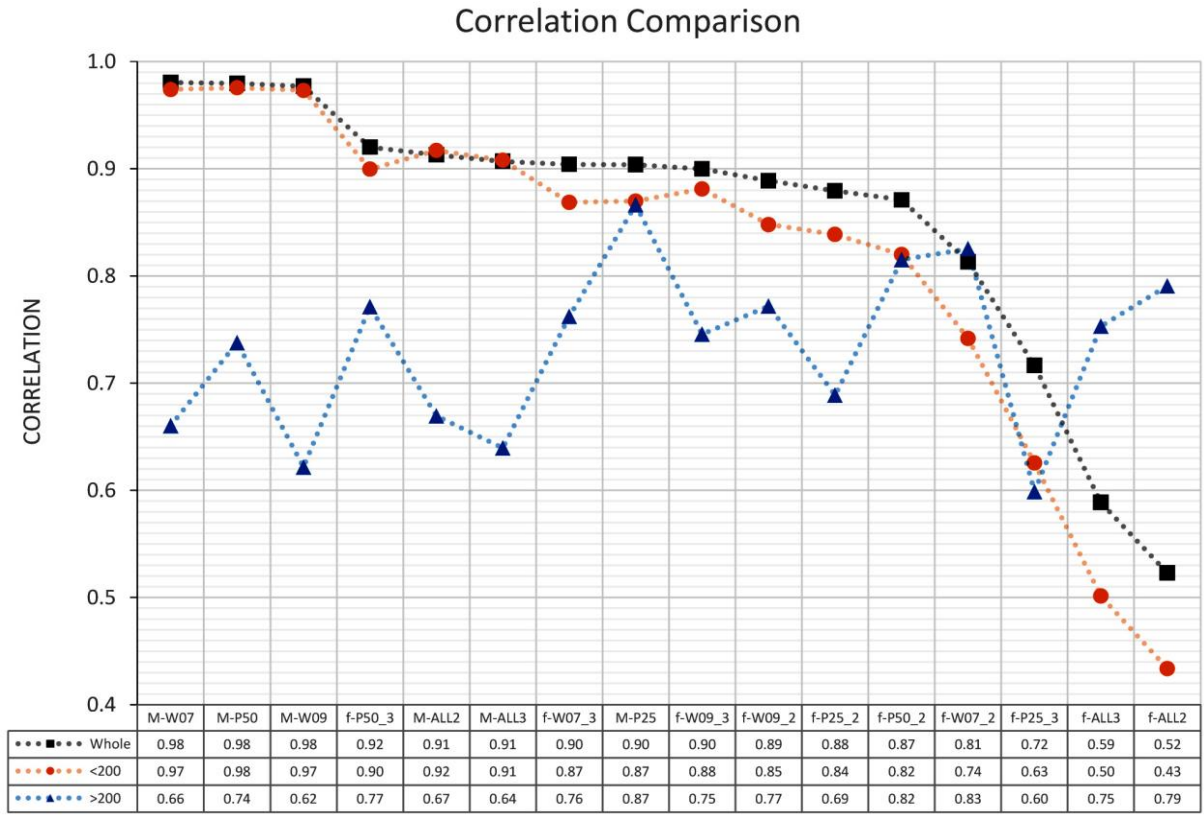

**Figure 16.** Correlation comparison (Eq. 1) between filtered and unfiltered Moon Zoo crater data against expert count. Highest positive correlation is M-P50 with 0.98, lowest f-ALL2, with 0.43. Three filter/clustering combinations from the 'M-group' show excellent correlation (0.98) with control (RB) census for craters <200 m in diameter. When examined in isolation, >200 m craters in general fare worst in comparison, probably due to their overall much smaller representation and high users' errors and misinterpretations (see Figs. 10 and 13 background grey annotations). Ref. Table 4 for abbreviations key.

# TABLES

**Table 1.** NAC images and data used to constrain Apollo 17 LM geo-coordinates. Differences (Gr. dist.) relate to deviation (ground distance in metres) from most recent published location (Haase et al., 2012); "sample" and "line" refer to NAC line coordinates; "date" is the day of obtainment; "Res w_h" are width pixel and height pixel resolutions respectively'; "Inc. Angle" is illumination incident angle; "Flight Dir." 'Is' the N-S flying direction of the craft in relation to the lunar surface.

| SET | Sample | Line | Lat | Lon | Gr. dist. (m) | Date | Orbit # | Res w_h | Inc. Angle (°) | Flight Dir. |
|---|---|---|---|---|---|---|---|---|---|---|
| 180966380LE | 1825.81 | 12619.8 | 20.188 | 30.773 | 90 | 12/01/2012 | 11785 | 1.33_1.32 | 69.55 | +X |
| 183325253RE | 985.569 | 12823.3 | 20.188 | 30.771 | 81 | 08/02/2012 | 12115 | 1.32_1.32 | 43.89 | +X |
| 190394800RE | 2211.86 | 14088.8 | 20.190 | 30.774 | 88 | 30/04/2012 | 13104 | 1.32_0.76 | 45.77 | -X |
| 165645700RE | 858.126 | 43266.7 | 20.190 | 30.773 | 42 | 18/07/2011 | 9545 | 0.41_0.55 | 69.27 | -X |
| *Davies and Colvin, 2000* | | | 20.191 | 30.772 | 4 | | | | | |
| **Haase et al., 2012** | | | **20.191** | **30.772** | | | | | | |
| 134985003RE | 452.594 | 13899.2 | 20.191 | 30.770 | 40 | 28/07/2010 | 5026 | 0.49_0.55 | 63.81 | -X |
| 106690695RE | 3264.29 | 24483.7 | 20.194 | 30.771 | 96 | 04/09/2009 | 865 | 1.65_1.47 | 36.60 | -X |
| **104318871RE** | **2518.72** | **21336.3** | **20.195** | **30.769** | **150** | **07/08/2009** | **534** | **1.39_1.40** | **57.63** | **-X** |
| **104311715RE** | **2203.06** | **20634.6** | **20.195** | **30.768** | **171** | **07/08/2009** | **533** | **1.45_1.40** | **56.73** | **-X** |
| 101956806LE | 1337.26 | 21182.1 | 20.200 | 30.770 | 294 | 11/07/2009 | 204 | 1.37_1.39 | 81.43 | -X |
| 101949648RE | 1881.16 | 19653.6 | 20.201 | 30.767 | 344 | 11/07/2009 | 203 | 1.47_1.39 | 80.47 | -X |
| 101963963RE | 1473.09 | 21008.4 | 20.203 | 30.769 | 384 | 11/07/2009 | 205 | 1.41_1.39 | 82.34 | -X |

**Table 2.** MoonZoo citizen science crater annotations analysis – binned number of entries per user. Bin sizes where arbitrarily chosen to highlight trends. Right table shows key statistical data relating to users' crater annotations, i.e., maximum number of craters by single user is 1453, with a median of 5 for the 9321 citizen scientists. These figures relate solely to the four NAC images used in this work and they might represent just a small fraction of the overall output of each user.

| Number of entries per user (binned) | Frequency | % | Part. % sum |
|---|---|---|---|
| 1 | 1561 | 16.75 | |
| 10 | 5220 | 56.00 | **72.75** |
| 20 | 1233 | 13.23 | |
| 30 | 472 | 5.06 | **91.04** |
| 50 | 383 | 4.11 | |
| 70 | 160 | 1.72 | |
| 100 | 107 | 1.15 | |
| 200 | 131 | 1.41 | **99.42** |
| 300 | 29 | 0.31 | |
| 400 | 7 | 0.08 | |
| 500 | 4 | 0.04 | |
| 600 | 5 | 0.05 | |
| 700 | 3 | 0.03 | |
| 800 | 0 | | |
| 900 | 2 | 0.02 | |
| 1000 | 2 | 0.02 | |
| 1100 | 0 | | |
| 1200 | 1 | 0.01 | |
| 1300 | 0 | | |
| 1400 | 0 | | |
| 1500 | 1 | 0.01 | |

| PARAMETER (crater counts) | |
|---|---|
| **Mean** | 14 |
| Standard Error | 0.4 |
| Median | 5 |
| Mode | 1 |
| Standard Deviation | 42 |
| Sample Variance | 1782 |
| Kurtosis | 340 |
| Skewness | 15 |
| Minimum | 1 |
| Maximum | 1453 |
| Sum | 129479 |
| Count | 9321 |
| Confidence Level (95.0%) | 0.86 |

**Table 3.** Permutations of users filtering (pre-clustering), P25, P50, W07, W09 (or not = ALL) and clustering minimum number of annotations per crater thresholds (none, _2, or _3). Each permutation produces a different number of final craters from multiple Moon Zoo annotations (post-threshold). Last column shows discrepancies expressed in percentile against expert count (2,602 craters, RB).

| Filtering-Clustering (F-C) combinations | Post-F-C number of craters | Reduction from Pre-clustered set (%) | Comparison with RB count (%) |
|---|---|---|---|
| Expert (RB) count | 2,602 | | |
| **f-ALL_2** | 7,636 | 85.2 | 193 |
| **f-ALL_3** | 4,685 | 90.9 | 80 |
| **f-P50_2** | 1,694 | 96.7 | -35 |
| **f-P50_3** | 750 | 98.5 | -71 |
| **f-P25_2** | 530 | 99.0 | -80 |
| **f-P25_3** | 192 | 99.6 | -93 |
| **f-W07_2** | 2,134 | 95.9 | -18 |
| **f-W07_3** | 996 | 98.1 | -62 |
| **f-W09_2** | 1,226 | 97.6 | -53 |
| **f-W09_3** | 484 | 99.1 | -81 |
| **M-ALL_2** | 4,067 | 92.1 | 56 |
| **M-ALL_3** | 2,247 | 95.6 | -14 |
| **M-P50** | 3,260 | 93.7 | 25 |
| **M-P25** | 1,522 | 97.1 | -42 |
| **M-W07** | 4,027 | 92.2 | 55 |
| **M-W09** | 3,029 | 94.1 | 16 |

**Table 4.** Key to the abbreviations used in this work. '*f*-P50_3' means that crater data from users using more than 50% of the time (minimum) default notations were discarded ('-P50'). Data were then clustered using the *fastcluster* approach ('*f*-') with three craters notation set as a minimum ('_3') to generate a coalesced crater entry, which is an average of the three craters positions and radii.

| Users's data Filtering out | Abbreviation |
|---|---|
| >50 % def. size | -P50 |
| >25 % def. size | -P25 |
| 'looser' weighing | -W07 |
| 'stricter' weighing | -W09 |
| **Clustering methods** | |
| *fastcluster* | *f*- |
| Manchester | M- |
| **Min. crater annotations** | |
| at least 2 | _2 |
| at least 3 | _3 |